\documentstyle[aps,epsf]{revtex}
\begin{document}
\draft
\twocolumn[\hsize\textwidth\columnwidth\hsize\csname @twocolumnfalse\endcsname
\title{Backscattering-Induced Crossover in Deterministic Diffusion}
\author{R. Klages$^{(1,2)}$\cite{em}, J.R. Dorfman$^{(2)}$}
\address{$^{(1)}$Institut f\"ur Theoretische Physik, Technische
Universit\"at Berlin, Sekr.\ PN 7-1,\\ Hardenbergstr.\ 36, D-10623
Berlin, Germany\\
$^{(2)}$Institute for Physical Science and Technology and
Department of Physics,\\ 
University of Maryland, College Park, MD 20742, USA}
\date{\today}
\maketitle
\begin{abstract}
We study diffusion in a one-dimensional periodic array of scatterers
modeled by a simple map. The chaotic scattering process for this map
can be changed by a control parameter and exhibits the dynamics of a
crisis in chaotic scattering. We show that the strong backscattering
associated with the crisis mechanism induces a crossover which leads
to different asymptotic laws for the parameter-dependent diffusion
coefficient. These laws are obtained from exact diffusion coefficient
results and are supported by simple random walk models. We argue that
the main physical feature of the crossover should be present in many
other dynamical systems with non-equilibrium transport.
\end{abstract}
]
One of the basic mechanisms in the theory of chaotic dynamical systems 
are so-called crisis events, where the asymptotic dynamics of the
system changes drastically with respect to the variation of a control
parameter \cite{COY82,Ott,ASY96}. Recently, it was discovered that
similar events occur in simple chaotic scattering systems when the
scattering rules are varied. This phenomenon has been denoted as a
crisis in chaotic scattering \cite{LaGr94}. On the other hand, over
the past few years a considerable literature has been developed in
which the origin of transport in non-equilibrium statistical mechanics
has been related to the characteristics of chaotic scattering
processes \cite{tacs}. One problem studied was deterministic diffusion
in simple one-dimensional maps \cite{GF1,SFK,mdd}, where for some
examples parameter-dependent diffusion coefficients have been computed
by taking the complete equations of motion of the dynamical systems
into account \cite{RKD,Diss}. Similar one-dimensional maps have been
proposed in Ref.\ \cite{LaGr94} as simple models for a crisis in
chaotic scattering. Thus, the aim of this Letter is to investigate
whether a crisis in chaotic scattering has an impact on deterministic
diffusion in simple one-dimensional maps.

\begin{figure}[b]
\epsfxsize=8cm
\centerline{\epsfbox{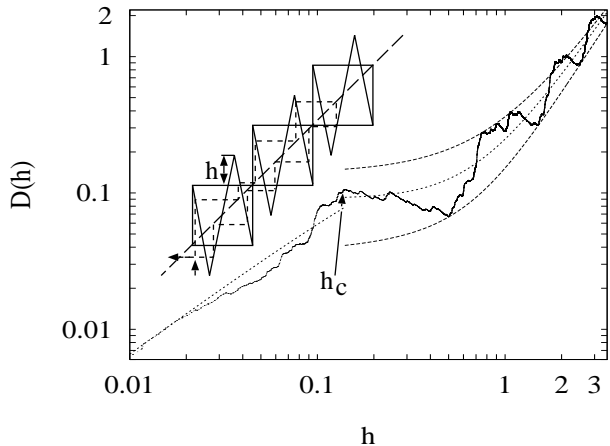}}
\vspace*{0.3cm} 
\caption{Double logarithmic plot of the diffusion coefficient $D(h)$
with respect to the height $h$ for the sawtooth map shown in the
figure. The graph is based on 38,889 single data points. Two random
walk solutions (dotted lines) and two curves which approximately give
the boundaries of the oscillations of $D(h)$ for values of $h$ above
the crisis point $h_c$ (dashed lines) are included.}
\label{f1}
\end{figure}

In the following, we consider discrete one-dimensional piecewise
linear chaotic maps with uniform slope, $x_{n+1}=M_h(x_n)$, where $h$
is a control parameter, and $x_n$ is the position of a point particle
at the discrete time $n$. $M_h(x)$ is continued periodically beyond
the interval $[0,1]$ onto the real line by a lift of degree one,
$M_h(x+1)=M_h(x)+1$. We assume that $M_h(x)$ is anti-symmetric with
respect to $x=0$, $M_h(x)=-M_h(-x)$, i.e., that there is no drift
imposed on a point particle. As an example, we consider the sawtooth
map sketched in Fig.\ \ref{f1}. It was chosen as a periodic
continuation of the map studied in Ref.\ \cite{LaGr94} which exhibits
a crisis in chaotic scattering. The control parameter is here the
height $h$ of the map which is related to the absolute value of the
slope $a$ by $h=(a-3)/4$. The diffusive properties of similar maps
have been studied in Refs.\ \cite{GF1,mdd}. For this sawtooth map the
parameter-dependent diffusion coefficient has been computed by solving
the Frobenius-Perron equation of the dynamical system \cite{Ott},
\begin{equation} 
\rho_{n+1}(x) = \int dy \; \rho_n(y) \; \delta(x-M_h(y)) \quad , 
\label{eq:fp}
\end{equation}
where $\rho_n(x)$ is the probability density for points on the real
line, and $M_h(y)$ is the map under consideration. There exists a dense
set of parameter values $h$ for which one can construct Markov partitions
for the map, and for each of these parameter values Eq.\ (\ref{eq:fp})
can be written as a matrix equation \cite{RKD,Diss},
\begin{equation}
\mbox{\boldmath $\rho$}_{n+1}=(1/|a|) \, T \, \mbox{\boldmath
$\rho$}_n \quad . \label{eq:fpm}
\end{equation}
$\mbox{\boldmath $\rho$}_n$ represents a column vector of the
probability densities in each part of the Markov partition at time
$n$, and $T$ is a topological transition matrix which can be obtained
from the Markov partition. However, instead of solving the eigenvalue
problem of $T$ \cite{RKD}, here solutions for the probability density
vector $\mbox{\boldmath $\rho$}_n$ have been obtained by iterating
Eq.\ (\ref{eq:fpm}), 
\begin{equation}
\mbox{\boldmath $\rho$}_{n+1}=(1/|a|^n) T^n
\mbox{\boldmath $\rho$}_0 \quad . 
\end{equation}
Starting with any probability density vector $\mbox{\boldmath
$\rho$}_0$ this iteration method enables us to compute the exact
time-dependent probability density $\rho_n$ at any time step $n$ and
all other dynamical quantities based on probability density averages
for maps of the type of $M_h(x)$ \cite{Diss,tbp,itm}. In particular,
it provides an efficient way to calculate diffusion coefficients by
employing an Einstein formula,
\begin{equation}
D(h)=\lim_{n\to\infty}\frac{1}{2n}\int dx \,\rho_n(x) x^2 \quad ,
\label{eq:dke}
\end{equation}
where the integral is the second moment of the time-dependent
probability density \cite{dcm}.

Fig.\ \ref{f1} shows a log-log plot of the diffusion coefficient as a
function of $h$ up to $h=3.5$. Included are four curves which describe
the coarse-grained behavior of the exact results. For integer values
of the height the diffusion coefficient can be computed analytically
by applying the eigenvalue method of Ref.\ \cite{RKD}, and we get
\begin{equation}
D(h)=\frac{2h^3+3h^2+h}{12h+9} \to \frac{h^2}{6}\; (h\to \infty) \; ,
\; h \,\in\, N\; . \label{eq:dke1} 
\end{equation}
The two {\em dashed} curves give approximate limits for the
oscillations of the exact diffusion coefficient in the range
$h>h_c$. They are obtained by fitting the diffusion coefficient with
the functional form of Eq.\ (\ref{eq:dke1}) at $h=(2k+1)/2$ and
$h=(4k+3)/4\, ,\,k\,\in N_0$, for the upper and lower curve,
respectively. The two {\em dotted} curves show two simple random walk
approximations. For large heights the distance a point particle
travels at one time step by moving from one unit interval to another
is taken into account exactly \cite{Ott}, and we get
\begin{equation} 
D_{rw1}(h)= \int_0^{1/2}dx (M_h(x)-x)^2\to \frac{h^2}{6}
\label{eq:drw2s} \; (h\to \infty)\; ,
\end{equation} 
which gives the dotted line plotted for $h>h_c$. For small heights the
absolute value of the distance is approximated to either zero or one,
depending on whether the particle remains on a unit interval or moves
from one unit interval to the next \cite{SFK}. This leads to
\begin{equation} 
D_{rw2}(h)=\frac{2h}{4h+3}\to \frac{2}{3}h \quad
(h\to 0) \quad .  \label{eq:drw1s} 
\end{equation}
These approximations indicate three different regions of
coarse-grained behavior for the exact diffusion coefficient: The first
one is a simple initial region, where the diffusion coefficient behaves
linearly for small heights. For $h\ge h_c$ it decreases slightly on
increasing the height. Finally, for $h\ge 0.5$ it starts to grow
quadratically in the height, but with strong oscillations on a fine
scale. The transition between the two different types of asymptotic
coarse-grained behavior, which occurs in the intermediate region of
$h_c\le h\le 0.5$, can be understood by referring 

\noindent to the action of certain microscopic scattering mechanisms
of the map, which are closely connected to the dynamics of a crisis in
chaotic scattering.

\begin{figure}[b]
\epsfxsize=8cm
\centerline{\epsfbox{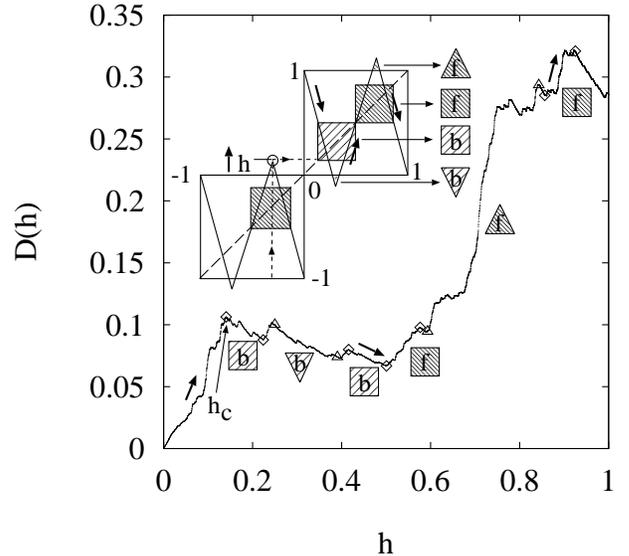}}
\vspace*{0.3cm} 
\caption{Chaotic scattering in the sawtooth map and its connection to
the behavior of the diffusion coefficient $D(h)$. Certain microscopic
scattering mechanisms of the map are identified by shaded squares and
triangles. The same symbols are shown along the $D(h)$ curve, where
they indicate the impact of the respective scattering regions on the
diffusion coefficient. The graph consists of 10,268 single data
points.}
\label{f2}
\end{figure}

These scattering mechanisms are introduced in Fig.\ \ref{f2}, where
certain regions of the map have been distinguished by shaded squares
and triangles: The triangles refer to parts where points of one unit
interval get mapped from that particular interval into another unit
interval. Additionally, if points enter a square they preferably move
into the triangular escape region above or below the respective square
after some iterations. These squares are identical to the squares of
an analogous scattering model, where they provide the fundamental
mechanism for a crisis in chaotic scattering \cite{LaGr94}. The
abbreviations {\em f} (``forward'') and {\em b} (`` backward'') in
these scattering regions refer to the dynamics of the critical point
of the map, which is indicated by a small circle. Its first iteration
is shown by the dashed line with the arrows. At its second iteration,
and by increasing the height $h$ continuously up from zero, the orbit
of the critical point, denoted as the {\em critical orbit} in the
following, travels along the graph of the map in the next right box
from the upper left to the lower right, as indicated by bold black
arrows. This way, the critical orbit explores all the different
scattering regions of the map as $h$ is increased from zero. If it
hits a region labeled by a {\em b} it is in a position to get {\em
backscattered} into the box to the left. {\em Vice versa}, if the
orbit enters an {\em f} region it is in a preferable position to move
further {\em forward} to the next box to the right. The critical point
indicated in Fig.\ \ref{f2} is part of a forward scattering
region. Note that there is a dense set of points around the critical 
point 

\noindent which exhibits the same dynamics, at least for
the first few iterations. An event dynamically analogous to the one of
a crisis in chaotic scattering now occurs at the {\em crisis point},
defined by the parameter value of the height $h_c$ for which the
critical orbit of the forward scattering region hits the boundary of a
backward scattering region in the next right box after one iteration
for the first time. This case is illustrated in Fig.\ \ref{f2} and,
for the map shown, is determined by
$h_c=(\sqrt{17}-3)/8\simeq0.1404$. We remark that this process is {\em
topologically} not identical to a crisis in chaotic scattering in the
sense that a crisis is generated by the merging of two formerly
isolated invariant sets in the phase space. Nevertheless, we argue
that {\em dynamically} this process provides the same characteristics
as a crisis in chaotic scattering, especially the onset of strong
backscattering \cite{LaGr94}.

The squares and the triangles along the diffusion coefficient curve
now refer to parameter regions where the critical orbit gets mapped
into the respective scattering regions. The different symbols on the
curve denote boundary points where the critical orbit enters, or
leaves, these regions. The diffusion coefficient clearly decreases
globally if the critical orbit enters a backscattering region, and it
increases globally if it exhibits forward scattering. Hence, the
different microscopic scattering mechanisms defined above are
connected to regions in the macroscopic diffusion coefficient which
exhibit different parameter-dependent behavior. Most importantly, the
crisis point $h_c$ corresponds to the first strong local maximum of
the curve, as shown in Fig.\ \ref{f2}. Magnifications reveal that the
fine structure of the curve changes significantly just below and above
the crisis value, and that the curve is fractal \cite{RKD}. This can be
understood in detail by refining the procedure explained above and is
reported elsewhere \cite{Diss,tbp}. Moreover, in Fig.\ \ref{f1} the
crisis point $h_c$ separates the regions which are described by the
two different random walk models and which show significantly
different coarse-grained behavior.

We conclude that the onset of strong backscattering affects the
diffusion coefficient of the sawtooth map not only on a fine scale,
but also on a coarse-grained scale. This phenomenon may be understood
as a {\em backscattering-induced crossover in deterministic
diffusion}. However, such a dynamical event must eventually take place
in any map of the type of $M_h(x)$ at a certain parameter value,
independently of its special functional form and independently of the
special topological characteristics of a crisis in chaotic
scattering. This can be checked by identifying the forward and
backward scattering regions of a map and applying the definition of
the crisis point given above. As an example, the piecewise linear,
discontinuous, non-sawtooth map studied in Ref.\ \cite{RKD} has been
analyzed. It turns out that the respective crisis point of the map
again corresponds to the first strong local maximum of the diffusion
coefficient curve, and that again the crisis point is related to a
change between two different laws for the parameter-dependent
diffusion coefficient on a coarse-grained scale: As before, the
diffusion coefficient grows linearly for small values of the height,
although with a slope different from that obtained for the sawtooth
map. It lacks a broad crossover region right above the crisis point,
but as in case of the sawtooth map it increases quadratically with a
factor of $1/6$ in the limit of large heights. Analogous results for
asymptotic diffusion coefficients have been reported in Ref.\
\cite{GF1} for a variety of other maps of the type of $M_h(x)$,
although in this previous work diffusion coefficients could be
computed exactly only for special values of the height. This leads us
to conjecture that a backscattering-induced crossover, as described
above, is typical for diffusive maps of the type of
$M_h(x)$. Moreover, we suspect that for these maps the
parameter-dependent coarse-grained diffusion coefficient always
decreases linearly in the limit of small heights, and that it
increases quadratically in the limit of large heights with a universal
factor of $1/6$. Similar results concerning the asymptotic
coarse-grained behavior of diffusion coefficients may be found in
certain classes of two-dimensional maps, as, e.g., in standard and
sawtooth maps, where quadratic laws in the limit of large control
parameters have already been obtained \cite{Ott,tdm}.

We suppose that the main physical feature of the crossover, i.e., the
connection between the onset of strong microscopic backscattering and
a change in the behavior of macroscopic parameter-dependent transport
coefficients, is quite common not only in discrete one- and
two-dimensional models, but also in higher-dimensional Hamiltonian
systems: For example, in Ref.\ \cite{KnaNo} diffusion of an electron
in a two-dimensional crystal, modeled by Coulombic periodic
potentials, has been studied. For this class of dynamical systems an
energy threshold has been proved to exist above which the diffusion
coefficient increases with a power law in the particle energy, whereas
below this threshold no diffusion coefficient exists. In Ref.\
\cite{LaGr94} it has been found that related models exhibit a crisis
in chaotic scattering. Hence, we conjecture that the existence of this
energy threshold is linked to the dynamics of a crisis in chaotic
scattering as in case of the backscattering-induced crossover
discussed above. Furthermore, for the models with a crisis in chaotic
scattering the significance of orbiting collisions has been pointed
out, which indicate the onset of strong backscattering
\cite{LaGr94}. However, orbiting collisions have already been
discussed extensively for fluid systems at low densities, where
particles interact via Lennard-Jones potentials, and a qualitative
connection between the onset of these collisions and a small change in
the temperature-dependent behavior of transport coefficients has been
established \cite{Kla92}. Thus, physically a direct line may be drawn
from the dynamical origin of crises in dynamical systems over the
occurrence of certain microscopic chaotic scattering processes to a
specific coarse-grained behavior of transport coefficients, which may
be related to macroscopic dynamical crossover phenomena, or in certain
cases possibly even to dynamical phase transitions.

We conclude with a few remarks: (a) The sawtooth map under
consideration here was chosen as a diffusive version of a
one-dimensional dynamical system which exhibits a crisis in chaotic
scattering. A crossover in the parameter-dependent diffusion
coefficient of this map has been found which is generated by the
dynamical mechanism of this crisis event. It affects the diffusion
coefficient not only on a fine scale, but also on a coarse-grained
scale and leads to two different algebraic laws for the asymptotic
diffusion coefficient. (b) These different algebraic laws, i.e., a
linear decrease of the diffusion coefficient for small heights and a
quadratic increase with a factor of $1/6$ in the limit of large
height, are suspected to be universal for maps of the type of
$M_h(x)$. (c) The crossover is understood physically by relating the
onset of strong microscopic backscattering to a change in the behavior
of the macroscopic parameter-dependent transport coefficient. This
main physical feature is conjectured to be quite common in dynamical
systems which exhibit non-equilibrium transport.

Helpful discussions with Chr.\ Beck, A.\ Latz, E.\ Ott, E.\
Sch\"{o}ll, T.\ T\'{e}l, and J.\ Yorke are gratefully
acknowledged. R.K.\ is indebted to C.\ Grebogi for his continuing
interest and support of this work, and he wants to thank S.\ Hess, the
NaF\"oG Commission Berlin, the IPST, the Friends of the TU Berlin, and
the DFG for financial and other support. J.R.D.\ wishes to acknowledge
support from the National Science Foundation under grant
PHY-93-21312.


\begin{references}
\vspace*{-1cm}
\bibitem[*]{em}Electronic address: rkla@ipst.umd.edu
\bibitem{COY82} C.~Grebogi, E.~Ott, and J.A.~Yorke, Phys.\ Rev.\ Lett.\
{\bf 48}, 1507 (1982); Physica D {\bf 7}, 181 (1983); C.~Grebogi et al.,
Phys.\ Rev.\ A {\bf 36}, 5365 (1987)
\bibitem{Ott} E.~Ott, {\em Chaos in Dynamical Systems} (Cambridge
University Press, Cambridge, 1993)
\bibitem{ASY96} K.T. Alligood, T.~Sauer, J.A.~Yorke, {\em Chaos:
An introduction to dynamical systems} (preprint, 1996)
\bibitem{LaGr94} Y.-Ch.~Lai et al., Phys.\ Rev.\ Lett.\ {\bf 71}, 2212
(1993); Phys.\ Rev.\ E {\bf 49}, 3761 (1994)
\bibitem{tacs} See P.~Gaspard, J.R.~Dorfman, Phys.Rev.\ E {\bf 52},
3525 (1995), and references therein
\bibitem{GF1} H.~Fujisaka, S.~Grossmann, Z.\ Physik B {\bf 48}, 261
(1982); Phys.\ Rev.\ A {\bf 26}, 1179 (1982); Ch.-Ch.~Chen, Phys.\
Rev.\ 
\newpage
E {\bf 51}, 2815 (1995)
\bibitem{SFK} M.~Schell, S.~Fraser, R.~Kapral, Phys.\ Rev.\ A {\bf 26},
504 (1982)
\bibitem{mdd} S.~Grossmann, S.~Thomae, Phys.\ Lett.\ A {\bf 97},
(1983); T.~Geisel, S.~Thomae, Phys.\ Rev.\ Lett.\ {\bf 52}, 1936
(1984); R.~Artuso, Phys.\ Lett.\ A {\bf 160}, 528 (1991); Physica D
{\bf 76}, 1 (1994); R.~Artuso, G.~Casati, R.~Lombardi, Phys.\ Rev.\
Lett.\ {\bf 71}, 62, (1993); H.-Ch.~Tseng, Phys.\ Lett.\ A {\bf 195},
74 (1994)
\bibitem{RKD} R.~Klages, J.R.~Dorfman, Phys.\ Rev.\ Lett.\ {\bf 74},
387 (1995)
\bibitem{Diss} R.~Klages, {\em Deterministic diffusion in
One-Dimensional Chaotic Dynamical Systems}, Dissertation, TU Berlin,
1995 (published in: Wissenschaft \& Technik Verlag, Berlin, 1996)
\bibitem{tbp} R.~Klages, J.R.~Dorfman (to be published)
\bibitem{itm}Applying the iteration method to large sets of parameter
values of the height, time-dependent Gaussian probability densities
with strong periodic fine structures have been obtained for the
sawtooth map and for the piecewise linear map of Ref.\ \cite{RKD}. 
Moreover, the kurtosis, time-dependent diffusion coefficients, and
velocity autocorrelation functions have been computed. All these
quantities show the characteristics of a simple statistical diffusion
process on a coarse-grained scale, i.e., they increase linearly, or
decrease exponentially, respectively, but with certain oscillations on
a fine scale \cite{Diss,tbp}.
\bibitem{dcm}The iteration method leads to values for $D(h)$ which are
precise up to an order of $10^{-7}$ after maximally 15 iteration
steps. Therefore, no error bars appear for the results presented in
Figs.\ \ref{f1} and \ref{f2}. Another approach to compute diffusion
coefficients for maps of the type of $M_h(x)$ is based on a Green-Kubo
formula and relates the parameter-dependent diffusion coefficient to a
class of fractal functions which is defined by certain functional
equations \cite{Diss,tbp}. An even more efficient method has been
developed by J.~Groeneveld (to be published).
\bibitem{tdm} see I.~Dana, N.W.~Murray, I.C.~Percival, Phys.\ Rev.\
Lett.\ {\bf 62}, 233 (1989); J.D.~Meiss, Rev.\ Mod.\ Phys.\ {\bf 64},
795 (1992), and further references therein
\bibitem{KnaNo} A.~Knauf, Commun.\ Math.\ Phys.\ {\bf 110}, 89 (1987);
B.~Nobbe, J.\ Stat.\ Phys.\ {\bf 78}, 1591 (1995); Diploma Thesis, TU
Berlin (1995, unpublished)
\bibitem{Kla92} R.~Klages, S.~Hess, W.~Loose, Verhandl.\ DPG (VI) {\bf
26}, 1002 (1991); R.~Klages, Diploma Thesis, TU Berlin (1992,
unpublished)
\end{references}
\end{document}